\documentclass{aa}

\usepackage{txfonts}

\DeclareSymbolFont{AMSb}{U}{msb}{m}{n}
\DeclareSymbolFontAlphabet{\Bbb}{AMSb}

\ifx\pdfoutput\thisisundefined
  \newcommand{\dofig}{P}
\else
  \newcommand{\dofig}{D}
\fi

\if\dofig D
  \usepackage[pdftex]{graphicx}
  \newcommand{\putfig}[2]{\includegraphics[height=#2]{#1.pdf}}
\else
  \usepackage[dvips]{graphicx}
  \newcommand{\putfig}[2]{\includegraphics[height=#2]{#1.eps}}
\fi

\newcommand{\sub}[1]{_\mathrm{#1}}

\newcommand{\keyw}[1]{\mbox{{\tt #1}}}

\newcommand{\keyi}[2]{\mbox{{\tt #1\hspace{1pt}}{$#2$}\/}}
\newcommand{\CTYPE}[1]{\keyi{CTYPE}{#1}}

\newcommand{\PVi}[1]{\mbox{{\tt PV\hspace{1pt}{$i$}\_{#1}{$a$}}\/}}

\newcommand{\keyv}[1]{\mbox{{\tt #1}}}

\begin{document}

\title{Mapping on the HEALPix grid}

\author{M. R. Calabretta\inst{1}}

\institute{Australia Telescope National Facility,
           PO Box 76, Epping, NSW 1710, Australia}

\offprints{M. Calabretta, \\
           \email{mcalabre@atnf.csiro.au}}

\date{Preprint submitted to astro-ph on 2004/12/23}

\abstract{The natural spherical projection associated with the Hierarchical
  Equal Area and isoLatitude Pixelisation, {\em HEALPix}, is described and
  shown to be one of an infinite class not previously documented in the
  cartographic literature.  Projection equations are derived for the class in
  general and it is shown that the HEALPix projection suggests a simple method
  (a) of storing, and (b) visualising data sampled on the grid of the HEALPix
  pixelisation, and also suggests an extension of the pixelisation that is
  better suited for these purposes.  Potentially useful properties of other
  members of the class are described.  Finally, the formalism is defined for
  representing any member of the class in the FITS data format.
}

\maketitle


\section{Introduction}
\label{sec:intro}

The Hierarchical Equal Area and isoLatitude Pixelisation, {\em HEALPix}
(G\'orski et al.\ \cite{kn:GHB}) offers a scheme for distributing $12 N^2
(N \in \Bbb{N})$ points as uniformly as possible over the surface of the unit
sphere subject to the constraint that the points lie on a relatively small
number ($4N-1$) of parallels of latitude and are equally spaced in longitude
on each of these parallels.  These properties were chosen to optimise
spherical harmonic analysis and other computations performed on the sphere.

In fact, HEALPix goes further than simply defining a distribution of points;
it also specifies the boundary between adjacent points and does so in such a
way that each occupies the same area.  Thus HEALPix is described as an
{\em equal area pixelisation}.  Pixels at the same absolute value of the
latitude have the same shape, though pixel shape differs between latitudes.
The boundaries for $N = 1$ define the 12 {\em base-resolution pixels} and
higher-order pixelisations are defined by their regular subdivision.  Note,
however, that although they are four-sided, HEALPix pixels are not spherical
quadrilaterals because their edges are not great circle arcs.

HEALPix was originally described purely with reference to the sphere, the data
itself being stored as a one-dimensional array in a FITS binary table (Cotton
et al.\ \cite{kn:CTP}) with either {\em ring} or {\em nested} organization,
the former being suited for spherical harmonic analysis and the latter for
nearest-neighbour searches.  For visualisation purposes the software that
implements HEALPix offers a choice of four conventional projection types
onto which HEALPix data may be regridded.

\begin{figure}
  \centerline{\putfig{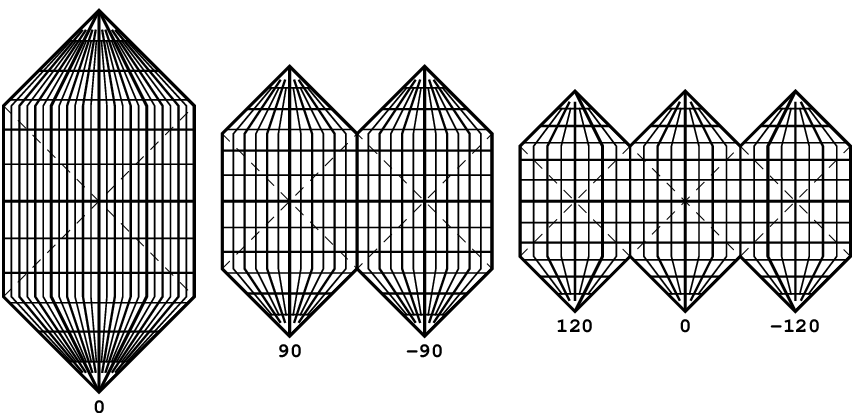}{117pt}}
  \vspace{5pt}
  \centerline{\putfig{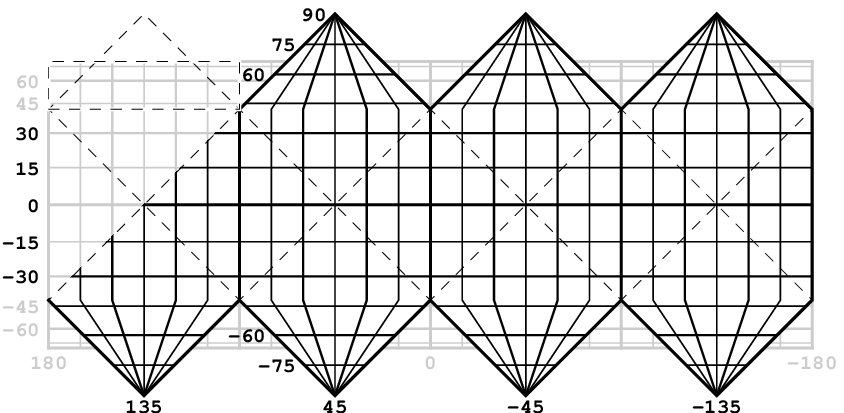}{117pt}}
  \caption[]{The HEALPix class of projections for $H = 1,2,3$ (top), and the
  nominative case with $H = 4$ (bottom) at twice the linear scale and with the
  top left-hand corner of the graticule ``cut away'' to reveal the underlying
  cylindrical equal-area projection in the equatorial region.  Facets are
  shown as dashed diamonds.}
  \label{fig:HPX}
\end{figure}

However the existence of the analytical mapping of each of the twelve
base-resolution pixels onto a $[0,1] \times [0,1]$ unit square was noted, and
in fact the HEALPix software uses this.  Roukema \& Lew (\cite{kn:RL}) have
recently provided a mathematical derivation of the equations and present a
diagram showing a projection of the whole sphere (hereinafter the
{\em HEALPix projection}) in which the base-resolution pixels, and
consequently the pixels of all higher-order pixelisations, are projected as
diamonds (i.e.\ squares rotated by 45\degr).  These equations may readily be
synthesised into those of an equal area projection of the whole sphere.

The HEALPix projection does not appear to have been documented previously in
cartography texts and could not be located in a web search; in particular, it
is absent from Snyder's (\cite{kn:Sny}) review of the history of cartography,
and also from Pearson (\cite{kn:Pea}).  Illinois State University's MicroCAM
website presents a catalogue of 320 map projections produced by a member of
the International Cartographic Association's Commission on Map Projections
(Anderson, \cite{kn:And}).  None bear even a superficial resemblence; the
equal-area quad-cube may be disected and rearranged to produce something with
a similar boundary but it is a distinctly different projection.  The stated
intention of this website is to present as complete a collection as possible
of historical, published map projections.

This work will show that the HEALPix projection is one of the more important
members of an infinite class of projections parameterised by $H \in \Bbb{N}$
and will derive the projection equations for the class.  In particular, the
HEALPix projection (i.e.\ with $H = 4$) suggests a simple way of storing
HEALPix data on a two-dimensional square grid as used in conventional imaging
and mapping, and also suggests an extension of the HEALPix pixelisation that
is better suited to this.  The HEALPix projections with $H = 3$, and $6$ are
also shown to be special, and their properties will be discussed.

The related issues of representing celestial coordinates in the HEALPix
projection are also considered in relation to image data storage in FITS
(Hanisch et al.\ \cite{kn:NOST}).


\section{The HEALPix projections}
\label{sec:HEALProj}

Firstly in this section we derive the projection equations for the infinite
class of HEALPix projections.  This leads naturally to the subject of mapping
HEALPix data on a regular grid and extensions of the HEALPix pixelisation that
arise.

\subsection{HEALPix derivation}
\label{sec:Deriv}

Figure~\ref{fig:HPX} shows the first four members ($H = 1, \ldots, 4$) of the
HEALPix projections.  They may be described as interrupted, equal area,
pseudo-cylindrical projections whose defining characteristics are
\begin{enumerate}
\item
  They are equi-areal; regions with equal areas on the sphere have equal areas
  in the plane of projection.
\item
  Parallels of latitude are projected as horizontal straight lines
  (interrupted in the polar regions) whence
  ${\textstyle \partial \vary}/{\textstyle \partial\phi} = 0$.
\item
  Parallels are uniformly divided (apart from interruptions).
\item
  The interruptions are defined by stacking equal-area diamonds (hereinafter
  {\em facets}) as shown in Fig.~\ref{fig:HPX}.  The facet that straddles
  $\pm 180\degr$ is split into halves in the graticule.
\end{enumerate}

\subsubsection{Transition latitude, $\theta_\times$}
In deriving the projection equations, note firstly that for any $H$ the total
area occupied by the half-facets in the north polar region is always $1/6$ of
the total area.  Since the projections are equi-areal, we equate the area of a
spherical cap on the unit sphere, $A = 2 \pi (1 - \sin\theta)$, with the
corresponding fraction of the total area, $4\pi/6$, to obtain the transition
latitude, $\theta_\times$, which is independent of $H$:
\begin{equation}
  \theta_\times = \sin^{-1} (2/3) \approx 41\fdg8103 .
\end{equation}

\subsubsection{Equatorial region}
The equatorial region, where $|\theta| \le \theta_\times$, is clearly a
cylindrical equal-area projection, i.e.\ $(x, \vary) = (\phi,
\alpha \sin\theta)$, where $\alpha$ is a constant determined by the
requirement that $\theta_\times$ be projected at the vertex of a square,
i.e.\ $\vary_\times = \pi/H = \alpha \sin\theta_\times$, whence
\begin{eqnarray}
      x & = & \phi , \\
  \vary & = & \frac{3\pi}{2H} \sin\theta .  \label{eq:yequ}
\end{eqnarray}

Because ${\textstyle \partial \vary}/{\textstyle \partial\phi} = 0$ for the
HEALPix projections the Jacobian reduces to
\begin{equation}
  J(\phi,\theta) =
        \frac{1}{\cos\theta}
        \frac{{\textstyle \partial x}}{{\textstyle \partial\phi}}
        \frac{{\textstyle \partial \vary}}{{\textstyle \partial\theta}} \cdot
        \label{eq:Jacob}
\end{equation}
This gives the ratio of an infinitesimal area in the plane of projection to
the corresponding area on the sphere.  In the equatorial regions it is
$3\pi/2H$, a constant, indicative of an equi-areal projection.  Note that the
Jacobian is inversely proportional to $H$, but the graticules in the top part
of Fig.~\ref{fig:HPX} were set to the same areal scale by scaling $x$ and
$\vary$ by $\sqrt{H}$.

\subsubsection{Polar regions}
In the polar regions the area north of $\theta \,(> \theta_\times)$ on the
unit sphere is $2 \pi (1 - \sin\theta)$ and, noting that the pole is projected
at $\vary = 2\pi/H$, in the plane of projection it is $H(2\pi/H - y)^2$.
Equating the ratio of these to the value of the Jacobian obtained for the
equatorial region and solving we obtain
\begin{equation}
  \vary = \pm \frac{\pi}{H} (2 - \sigma) , \label{eq:ypol}
\end{equation}
where the negative sign is taken for the south polar region, and
\begin{equation}
  \sigma = \sqrt{3(1-|\sin\theta\,|\,)} \label{eq:sigma}
\end{equation}
is the ratio of the distance of the pole from the parallel of $\theta$ to that
of the pole from the parallel of $\theta_\times$.

The equation for $x$ may be obtained readily by integrating
Eq.~(\ref{eq:Jacob}) with
${\textstyle \partial \vary}/{\textstyle \partial\theta}$ from
Eqs.~(\ref{eq:ypol}) \& (\ref{eq:sigma}) to produce $x = \sigma\phi + C$,
where $C$ is the constant of integration, thus indicating that the parallels
are uniformly divided.  It is instructive also to consider a geometrical
argument; the area of any triangle in the $(x,y)$ plane with its apex at the
pole and base along a given parallel of latitude depends only on the change in
$x$ between its base vertices and not on their location.  Since the projection
is equi-areal, $x$ must therefore vary linearly with $\phi$.

Applying the interruptions to the parallels (which in fact could be omitted or
done in other ways to produce different projection types) we have
\begin{equation}
  x = \phi\sub{c} + (\phi - \phi\sub{c}) \,\sigma , \\
\end{equation}
where
\begin{equation}
  \phi\sub{c} = \frac{\pi}{H} \left( 2 \left \lfloor
             \frac{(\phi + \pi) H}{2\pi}
           \right \rfloor + 1 \right) - \pi
\end{equation}
is the native longitude in the middle of a polar facet and
$\lfloor u \rfloor$\,, the {\em floor} function, gives the largest integer
$\le u$.

\subsubsection{Properties}
The most important feature of the HEALPix projections, indeed the underlying
rationale for the HEALPix pixelisation, is that they are equi-areal with
squared boundaries.  Thus they may be completely inscribed by diamonds of
equal area, the minimum number of which is $3H$ (the facets).  Each facet is
subject to further subdivision by $N^2$ smaller equal-sized diamonds that are
identified as {\em pixels}; their centre positions in $(\phi,\theta)$ may be
computed readily for any $(H,N)$ from the inverse of the above projection
equations as are cited in Sect.~\ref{sec:FITS}.  As explained by G\'orski et
al.\ (\cite{kn:GHB}), it is significant for spherical harmonic analysis that
the pixel centres lie on a relatively small number of parallels of latitude,
and that the facets may be subdivided in a hierarchical way.

Of course a pixelisation may be constructed similarly from a cylindrical
equal-area projection, but the HEALPix projections are much less distorted in
the polar regions than any such projection.  Consequently the HEALPix pixels
are much truer in shape when projected onto the sphere and their centres are
much more uniformly distributed.  As shown by the dashed lines in the
upper-left corner of Fig.~\ref{fig:HPX}, the equivalent portion of the
underlying cylindrical projection, being severely squashed at the pole, is
stretched upwards to twice its height and brought to a point; the pole itself
is thereby projected as $H$ points rather than a line.  However, this is
gained at the cost of introducing $H-1$ interruptions which should properly be
considered as extreme distortions, though of little consequence for the
pixelisation.

Evaluating the partial derivatives we find
\begin{eqnarray}
  \left( \frac{{\textstyle \partial x}}{{\textstyle \partial\phi}},
  \frac{{\textstyle \partial \vary}}{{\textstyle \partial\theta}} \right)
    & = & \left( 1, \frac{3\pi}{2H} \cos\theta \right)
          \textrm{\hspace{20pt} $\ldots$ equatorial}
	  \label{eq:pdequ} \\
  \left( \frac{{\textstyle \partial x}}{{\textstyle \partial\phi}},
  \frac{{\textstyle \partial \vary}}{{\textstyle \partial\theta}} \right)
    & = & \left( \sigma, \frac{3\pi}{2H} \frac{\cos\theta}{\sigma} \right)
          \textrm{\hspace{18pt} $\ldots$ polar}
	  \label{eq:pdpol}
\end{eqnarray}
which shows that in the polar regions $x$ is scaled directly, and $\vary$ is
scaled inversely by $\sigma(\theta)$ in order for the Jacobian to maintain
constancy.

To get some idea of the relative degree of distortion between members of the
class, consider first from Eqs.~(\ref{eq:yequ}) \& (\ref{eq:ypol}) that
$\vary$ scales as $1/H$ for any $\theta$, while $x$ is independent of $H$
except for defining the interruptions.  Hence the relative spacing of
parallels between the equator and poles is independent of $H$, as is evident
in Fig.~\ref{fig:HPX}, and the distortion is determined solely by the relative
$y:x$ scaling.

A spherical projection is {\em conformal} or {\em orthomorphic} (true shape)
at points where the meridians and parallels are orthogonal and equi-scaled.
The general equations of the cylindrical equal area projection expressed in
terms of the conformal or {\em standard} latitude, $\theta_\circ$, are
$(x,\vary) = (\phi,\sin\theta/\cos^2\theta_\circ)$ (e.g.\ see Sect. 5.2.2 of
Calabretta \& Greisen \cite{kn:CG}), whence from Eq.~(\ref{eq:yequ})
\begin{equation}
  \theta_\circ = \cos^{-1} \sqrt{\frac{2H}{3\pi}} \cdot
\end{equation}
For $H = 1,2,3,4$ this is $(62\fdg57, 49\fdg35, 37\fdg07, 22\fdg88)$; the
first two of these exceed $\theta_\times$ and hence are inadmissible, and
$\theta_\circ$ is undefined for higher values of $H$.  Since the latitude that
halves the area of the equatorial region is $\sin^{-1} (1/3) = 19\fdg47$,
independent of $H$, this suggests that the projection with $H = 4$ is the
least distorted in the equatorial regions.

\begin{figure}
  \centerline{\putfig{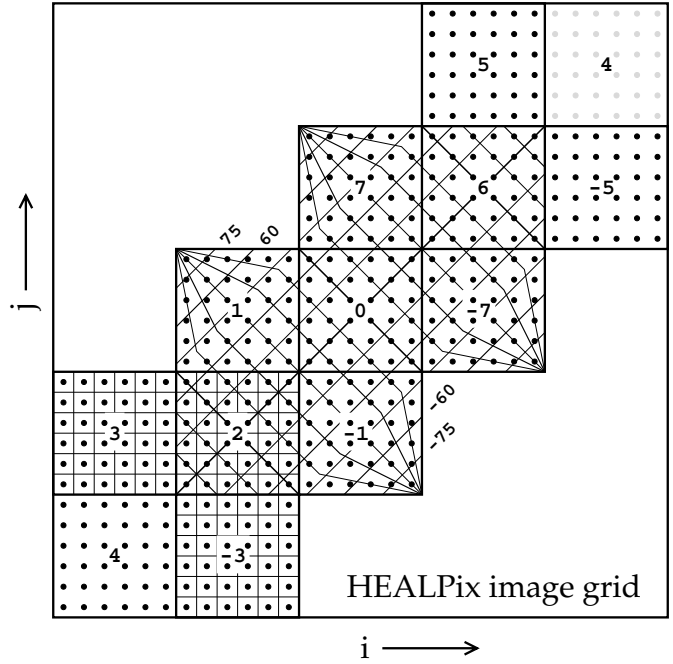}{250pt}}
  \caption[]{The HEALPix pixelisation for $N = 6$ on the HEALPix projection
  for $H = 4$ projected with a $45\degr$ rotation onto the mapping grid
  showing the twelve facets with a convenient numbering scheme.  The graticule
  of the HEALPix projection is shown in the seven facets adjacent to
  $(\phi,\theta) = (0,0)$, and those at lower left show the pixel boundaries
  for $N = 6$ as defined by the HEALPix pixelisation.}
  \label{fig:HPXgrid}
\end{figure}

Looking at it another way, the requirement for equiscaling in $x$ and $y$
where the meridians and parallels are orthogonal, i.e.\ everywhere in the
equatorial region, and along the centreline in the polar half-facets, is
\begin{equation}
  \frac{1}{\cos\theta}
  \frac{{\textstyle \partial x}}{{\textstyle \partial\phi}}
    = \frac{{\textstyle \partial \vary}}{{\textstyle \partial\theta}} \cdot
\end{equation}
Substituting Eqs.~(\ref{eq:pdequ}) and (\ref{eq:pdpol}) gives
\begin{eqnarray}
  H_\circ & = & \frac{3 \pi}{2} \cos^2\theta
                \textrm{\hspace{38pt} $\ldots$ equatorial}
		\label{eq:Hoequ}
		\\
  H_\circ & = & \frac{\pi}{2} (1 + |\sin\theta\,|)
                \textrm{\hspace{20pt} $\ldots$ polar, centreline}
		\label{eq:Hopol}
\end{eqnarray}
which gives $H_\circ = (4.7, 4.4, 3.5, 2.6, 2.7, 2.9, 3.1, 3.1)$ for
$\theta = (0, 15, 30, \theta_\times, 45, 60, 75, 90)$.  Thus $H = 4$ is a good
all-over compromise but for $|\theta| > 30\degr$, the latitude that halves the
area of the hemisphere, $H = 3$ would appear to be better on this basis.

\begin{figure}
  \centerline{\putfig{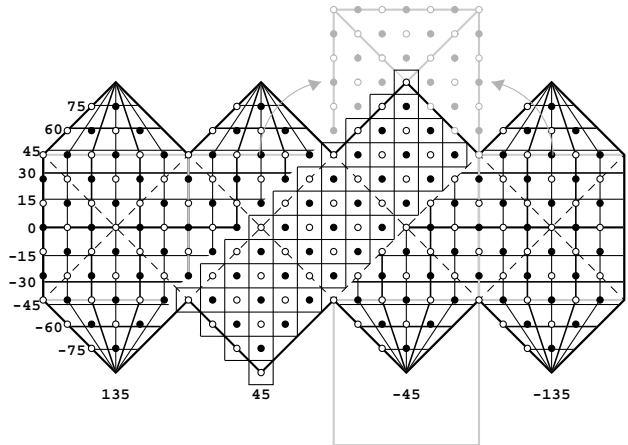}{167pt}}
  \caption[]{HEALPix double-pixelisation for $N = 3$ on the HEALPix projection
  with $H = 4$.  Filled circles define the regular grid, with interpolated
  pixels shown as open circles to the {\em left} of these.  Pixel boundaries
  are shown in the middle three facets, those of the two additional polar
  pixels contain a contribution from each of the four adjacent polar facets.
  Note the difference between the pixel locations in this figure compared to
  the $N = 6$ pixelisation in Fig.~\ref{fig:HPXgrid}.  Also illustrated is the
  boundary between the faces of the pseudo-quadcube layout of the HEALPix
  projection (grey).}
  \label{fig:HPXgrid2}
\end{figure}

The nature of the projective distortion in the region where meridians and
parallels are not orthogonal is more complicated.  In the polar regions the
projection of the facets onto the sphere (i.e. the base-resolution pixels)
meet at the pole at $360\degr/H$.  For $H = 4$ this is $90\degr$ which accords
with the angle in the plane of all HEALPix projections.  Thus it might seem
that $H = 4$ should be least distorted in the neighbourhood of the pole.
However, this argument is specious; on the sphere the angle between meridians
and parallels along the edges of the polar half-facets is always $90\degr$,
while in the plane of projection it is always $45\degr$.

These comments on distortion will be qualified in Sect.~\ref{sec:Other} for
members of the class for which additional $\vary$-scaling is applied.


\subsection{The HEALPix grid}
\label{sec:HPXgrid}

The base-resolution pixels of the HEALPix pixelisation are projected as
diamonds (squares rotated by $45\degr$) on the HEALPix projection with the
consequence that the pixel locations fall on a grid with diamond-symmetry.

However, Fig.~\ref{fig:HPXgrid} shows that the diamond grid may be converted
to the common square grid used in imaging via a trivial $45\degr$ rotation.
The resulting image plane is slightly less than half-filled (48\%) but this is
comparable to the figure of 50\% for quad-cube projections (O'Neill \&
Laubscher \cite{kn:OL}) which are commonly used in the same type of
application as HEALPix.  Moreover, being composed of square facets like the
quad-cubes, the HEALPix projection also admits the possibility of dissection
and storage on a third image axis, such as is implemented for the quad-cubes
via the \keyw{CUBEFACE} keyword in FITS (Calabretta \& Greisen (\cite{kn:CG}).
In this context Fig.~\ref{fig:HPXgrid2} shows how the facets may be
repartitioned into a configuration that resembles that of the quad-cube faces.
However, this resemblence is purely superficial because the ``cubeface'' edges
do not match those of a quad-cube projection on the sphere.

Facet number $4$ which straddles $\phi = \pm 180\degr$ may be treated in a
number of ways; it may be left split, or the halves may be reconnected in
either the lower-left or upper-right corner, or it may be replicated in both.

\subsubsection{HEALPix double-pixelisation}
The main drawback with the above technique for storing HEALPix imaging data is
that the image is presented at an unusual orientation.  However, this may be
solved via a simple extension to the HEALPix pixelisation.  Figure~3 shows the
HEALPix grid with a pixel interposed between every pair of pixels along the
parallels of latitude and additional pixels added at the two poles.  The total
number of pixels in the pixelisation is thereby increased from $12N^2$ to
$24N^2 + 2$ without affecting the special properties described by G\'orski
et al.\ (\cite{kn:GHB}), although requiring a slightly different method of
forming the hierarchy and indexing it.  Pixels that fall along the lines where
the polar half-facets meet are distorted in such a way as to ``zip'' the two
edges together but they still have equal area.


\subsection{Other pixelisations}
\label{sec:Other}

\begin{figure}
  \centerline{\putfig{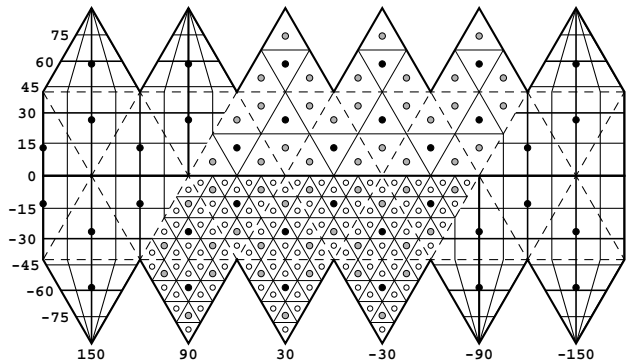}{134pt}}
  \caption[]{HEALPix projection for $H = 6$ scaled in $\vary$ by $\sqrt{3}$
  whereby the square facets become pairs of equilateral triangles.  These may
  then be subdivided in a hierarchical way; base-level pixels are shown to the
  left and right (black circles), the first level of subdivision is at mid-top
  (grey circles), the next at mid-bottom (open circles).}
  \label{fig:HPX6}
\end{figure}

Consider dividing the $360\degr$ of circumpolar latitude into integral
subdivisions.  Of the possible ways of doing this ($1 \times 360\degr,
2 \times 180\degr, 3 \times 120\degr, 4 \times 90\degr, 5 \times 72\degr,
6 \times 60\degr, \ldots$) only the divisions into 3, 4, and 6 correspond to
regular polyhedra.  The division into $4 \times 90\degr$ corresponds to the
familiar case of HEALPix with $H = 4$ with diamonds tesselated by diamonds.

\subsubsection{Triangular -- $H = 6$}
However, the division into $6 \times 60\degr$ suggests a different type of
pixelisation in which equilateral triangles are tesselated by equilateral
triangles.  This pixelisation may be defined by rescaling the HEALPix
projection with $H = 6$ by $\sqrt{3}$ in $\vary$ so that the half-facets
become equilateral triangles.  Such a linear scaling does not affect the
projection's equal area property.  What were previously half-facets may now be
identified with 36 new, triangular base-resolution pixels that may be
subdivided in a hierarchical way as for HEALPix, as depicted in
Fig.~\ref{fig:HPX6}.  It is interesting to note that this subdivision is
naturally hierarchical - the number of pixels varies exponentially as
$36 \times 4^{N-1}$ where $N$ is the hierarchy level.  In the $H = 4$
pixelisation the exponential hierarchy must be engineered by doubling $N$.

\begin{figure}
  \centerline{\putfig{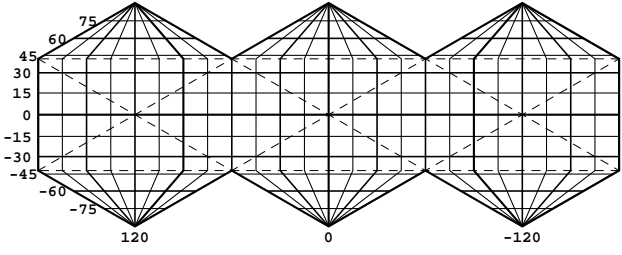}{92pt}}
  \caption[]{HEALPix projection for $H = 3$ scaled in $\vary$ by $1/\sqrt{3}$
  whereupon it becomes three consecutive hexagons.}
  \label{fig:HPX3}
\end{figure}

The conformal latitude computed for $H = 6$ with this extra $\vary$-scaling is
$\theta_\circ = 30\fdg98$, indicating that the projection becomes conformal at
the latitude that bisects the hemisphere by area.  Applying
Eqs.~(\ref{eq:Hoequ}) and (\ref{eq:Hopol}) with the extra scaling gives
$\sqrt{3} H_\circ = (8.2, 7.6, 6.1, 4.5, 4.6, 5.1, 5.3, 5.4)$ for $\theta =
(0, 15, 30, \theta_\times, 45, 60, 75, 90)$, again indicating less distortion
in the polar regions than $H = 4$.  It also does better in the polar zone away
from the centreline because the $60\degr$ angle along the edge of the polar
facets more closely matches the true angle of $90\degr$ on the sphere.
Overall, this pixelisation performs adequately at low latitudes and does
better than the $H = 4$ pixelisations at mid to high latitudes.

This rescaling of the $H = 6$ projection is reminiscent of Tegmark's
(\cite{kn:Teg}) icosahedral projection composed of 20 equilateral triangles;
the problem of indexing the subdivisions of its triangular facets was solved
in the implementation of the corresponding pixelisation.  In the present
context the iso-latitude property is still present but modified somewhat from
the diamond pixelisation of $H = 4$.  However, if the pixel centres are moved
up or down from the centroid by $\frac{1}{12}$ of the height of the
equilateral triangles to the point half-way between the base and apex then
they fall onto a regular grid sampled more frequently in $x$ than $y$.  This
provides some of the same benefits as the $H = 4$ double-pixelisation.

However, although the displacement is small, there is a possibility that it
will introduce statistical biases so the full consequences should be
investigated for a particular application.  These potential biases may be
minimised by making the pixel size sufficiently small, and the fact that the
bias between adjacent pairs of pixels is in opposite senses will tend to
cancel them over a region encompassing a sufficient number of pixels.  It
should also be remembered that although the pixel locations {\em appear} to be
at the centre of the pixel boundary in the projection of the diamond, square,
and triangular pixelisations this is very much an artifact of the distortions
inherent in the projection.  Because the $\vary$-coordinate varies
non-linearly with $\theta$, on the sphere they are actually biased to one side
of the pixel.  Hence some degree of bias is unavoidable.

\subsubsection{Hexagonal -- $H = 3$}
The division into $3 \times 120\degr$ suggests hexagonal base-resolution
pixels.  Although the familiar ``honeycomb'' structure shows that it is
possible to tile the plane with hexagons, nevertheless there is no bounded
tesselation of hexagons by hexagons.  That is, a hexagonal region may not be
cut out of a honeycomb tesselation without cutting the individual elements.
Thus it may seem surprising that a hexagonal pixelisation can be constructed
from the HEALPix projection for $H = 3$ with $\vary$ scaled by $1/\sqrt{3}$.
The boundary of this projection, as seen in Fig.~\ref{fig:HPX3}, is reduced to
that of three sequential hexagons.  This boundary is then used conceptually as
a ``pie-cutter'' on a honeycomb tesselation of the right scale.  Pixels that
are cut can be made whole again by borrowing from adjacent facets, much as the
square pixelisation in Fig.~\ref{fig:HPXgrid2} does.

Rescaling Eqs.~(\ref{eq:Hoequ}) and (\ref{eq:Hopol}) gives
$H_\circ / \sqrt{3} = (2.7, 2.5, $ $2.0, 1.5, 1.5, 1.7, 1.8, 1.8)$ for
$\theta = (0, 15, 30, \theta_\times, 45, 60, 75, 90)$.  Thus the rescaled
$H = 3$ projection does not achieve conformality at any latitude, it does well
close to the equator, but degrades at mid-latitudes.  In the polar regions the
$30\degr$ angle between meridians and parallels along the edge of the facets
is further from the ideal of $90\degr$ than the $45\degr$ for the unscaled
projections.


\subsection{\keyv{HPX}: HEALPix in FITS}
\label{sec:FITS}

In this section the HEALPix projections are described in the same terms as the
projections defined in Calabretta \& Greisen (\cite{kn:CG}).

HEALPix projections will be denoted in FITS with algorithm code \keyv{HPX} in
the \CTYPE{ia} keywords for the celestial axes.  As data storage has become
much less of an issue in recent years we do not consider it necessary to
create an analogue of the \keyw{CUBEFACE} keyword to cover \keyv{HPX}.
However, if HEALPix data is repackaged into the pseudo-quadcube layout shown
in Fig.~\ref{fig:HPXgrid2} the \keyw{CUBEFACE} storage mechanism is applicable
for $H = 4$ and will be treated properly by {\sc wcslib} (Calabretta,
\cite{kn:C2}).

Since the HEALPix projections are constructed with the origin of the native
coordinate system at the reference point, we set
\begin{equation}
   (\phi_0, \theta_0)\sub{HEALPix} = (0,0) .
\end{equation}
The projection equations and their inverses, re-expressed in the form required
by FITS, are now summarised formally.

In the equatorial zone where
$|\sin\theta\,| \le 2/3$:
\begin{eqnarray}
       x & = & \phi , \\
   \vary & = & \frac{270\degr}{H} \sin\theta ,
\end{eqnarray}
in the polar zones, where $|\sin\theta\,| > 2/3$:
\begin{eqnarray}
       x & = & \phi\sub{c} + (\phi - \phi\sub{c}) \,\sigma , \\
   \vary & = & \pm \frac{180\degr}{H} (2 - \sigma) ,
\end{eqnarray}
where the positive sign on $\vary$ is taken for $\theta > 0$, negative
otherwise, and
\begin{eqnarray}
       \sigma & = & \sqrt{3(1 - |\sin\theta\,|\,)} , \\
  \phi\sub{c} & = & \frac{180\degr}{H} \left( 2 \left \lfloor
                      \frac{(\phi + 180\degr) H}{360\degr}
                    \right \rfloor + 1 \right) - 180\degr .
\end{eqnarray}

These equations are readily invertible.  In the equatorial zone where
$|\,\vary\,| \le 180\degr/H$:
\begin{eqnarray}
    \phi & = & x , \\
  \theta & = & \sin^{-1} \left( \frac{\vary H}{270\degr} \right) ,
\end{eqnarray}
in the polar zones, where $|\,\vary\,| > 180\degr/H$:
\begin{eqnarray}
    \phi & = & x\sub{c} + (x - x\sub{c}) / \sigma , \\
  \theta & = & \pm \sin^{-1} \left( 1 - \frac{\sigma^2}{3} \right) ,
\end{eqnarray}
where the positive sign on $\theta$ is taken for $\vary > 0$, negative
otherwise, and
\begin{eqnarray}
    \sigma & = & 2 - \frac{|\,\vary H\,|}{180\degr} , \\
  x\sub{c} & = & \frac{180\degr}{H} \left( 2 \left \lfloor
                   \frac{(x + 180\degr) H}{360\degr}
                 \right \rfloor + 1 \right) - 180\degr .
\end{eqnarray}
where $x\sub{c}$ is the value of $x$ in the middle of a polar facet, as for
$\phi\sub{c}$.

FITS keyword \PVi{1} attached to {\em latitude} coordinate~$i$ will be used to
specify $H$ with default value 4.

\keyv{HPX} has been implemented in version 3.7 of {\sc wcslib} which is
distributed under a GNU Library Public License (GLPL).


\begin{acknowledgements}
The Australia Telescope is funded by the Commonwealth of Australia for
operation as a National Facility managed by CSIRO\@.
\end{acknowledgements}




\end{document}